\documentclass[aps,prl,preprint,superscriptaddress,showpacs]{revtex4}

\usepackage{graphicx}
\usepackage{amsmath}
\usepackage{amsfonts}
\usepackage{epsfig}  
\usepackage{color}

\begin{document}

\title{Thermal radiative near field transport between vanadium dioxide and silicon oxide across the metal insulator transition}

\author{F. Menges}
\affiliation{IBM Research - Zurich, S\"aumerstrasse 4, CH-8803 R\"uschlikon, Switzerland}

\author{M. Dittberner}
\affiliation{IBM Research - Zurich, S\"aumerstrasse 4, CH-8803 R\"uschlikon, Switzerland}
\affiliation{Photonics Laboratory, ETH Zurich, 8093 Zurich, Switzerland}

\author{L. Novotny}
\affiliation{Photonics Laboratory, ETH Zurich, 8093 Zurich, Switzerland}

\author{D. Passarello}
\affiliation{IBM Almaden Research Center, 650 Harry Road, San Jose, California 95120, USA}

\author{S. S. P. Parkin}
\affiliation{IBM Almaden Research Center, 650 Harry Road, San Jose, California 95120, USA}

\author{M. Spieser}
\affiliation{IBM Research - Zurich, S\"aumerstrasse 4, CH-8803 R\"uschlikon, Switzerland}

\author{H. Riel}
\affiliation{IBM Research - Zurich, S\"aumerstrasse 4, CH-8803 R\"uschlikon, Switzerland}

\author{B. Gotsmann}
\email{bgo@zurich.ibm.com}
\affiliation{IBM Research - Zurich, S\"aumerstrasse 4, CH-8803 R\"uschlikon, Switzerland}

\date{\today}


\begin{abstract}
The thermal radiative near field transport between vanadium dioxide and silicon oxide at submicron distances is expected to exhibit a strong dependence on the state of vanadium dioxide which undergoes a metal-insulator transition near room temperature.
We report the measurement of near field thermal transport between a heated silicon oxide micro-sphere and a vanadium dioxide thin film on a titanium oxide (rutile) substrate. The temperatures of the 15\,nm vanadium dioxide thin film varied to be below and above the metal-insulator-transition, the sphere temperatures were varied in a range between 100 and 200\,$^\circ$C. The measurements were performed using a vacuum-based scanning thermal microscope with a cantilevered resistive thermal sensor. We observe a thermal conductivity per unit area between the sphere and the film with a distance dependence following a power law trend and a conductance contrast larger than 2 for the two different phase states of the film.
\end{abstract}

\maketitle

At short distances, heat transport via thermal radiation can be enhanced with respect to the far field transport that follows the Stefan-Boltzmann law \cite{Cravalho1967}. This enhancement has been predicted and observed in many material systems \cite{Song2015}. At room temperature, near field effects become significant at distances below about one micron. The temperature dependence is weaker than for the far field case and scales at most with the square of the absolute temperature \cite{Basu2009}. Experimentally, near-field heat transport has been characterized by scanning thermal microscopes and dedicated MEMS structures \cite{Kittel:2005ws,Kim:jh}. Frequently, radiative thermal transport is measured between a sphere and a flat surface. Recent reviews by  Song et al. \cite{Song2015} and Jones et al. \cite{Jones2013} provide a comprehensive overview of experimental approaches to characterize radiative thermal transport.

A particularly strong enhancement of thermal radiative transport in the near field  has been predicted and observed for systems that exhibit strong interaction by coupling of surface phonon-polaritons (SPPs) \cite{Song2015}. Materials having SPPs are, for example, polar dielectrics like silicon dioxide (SiO$_2$), silicon carbide, or vanadium dioxide in its insulating state.
Among the materials supporting SPPs, vanadium dioxide (VO$_2$) is of particular interest as it undergoes a metal-insulator-transition (MIT) near room temperature. The MIT has been exploited for various applications including switches in electronic circuits \cite{Shukla2015, Kim2010} or optical modulators \cite{driscoll2009}. As VO$_2$ can support SPPs only in the insulating state, a strong modulation of thermal near field transport between VO$_2$ and other materials supporting SPPs is expected across the MIT. An increase in conductance of up to a factor of 100 at $10^{-8}$\,m distance between VO$_2$ and SiO$_2$ has been predicted \cite{Zwol2011}. However, the effect is expected to be reduced for thin (sub-micron) films of VO$_2$ \cite{Zwol2012, Yang2015}.

The interest to exploit this near-field switchability stems from the challenges to controlling thermal transport in solid state structures. In contrast to electronic or optical transport, methods of modulating or controlling heat conduction are scarce and inefficient, and the thermal conductivity difference between different solids varies only by few orders of magnitude (in contrast to tens of orders of magnitude for charge transport). Consequently, near field radiation interacting with VO$_2$ surfaces have been envisioned for thermal memory \cite{Kubytskyi2014, Dyakov2015}, logic \cite{Abdallah2014}, thermal diodes, and energy conversion devices \cite{Basu2009}.

Only a few attempts are reported to experimentally verify the thermal radiative switching of VO$_2$ across its MIT. A far field contrast larger than 2 has been observed \cite{Kats2013}. Near field measurements using SiO$_2$ glass spheres facing a thin film of VO$_2$ on a SiO$_2$ substrate showed a clear dependence on the phase state of VO$_2$, but the thermal conductance variation could not be quantified due to experimental constraints.

Here, we examine the near field thermal conductance between a SiO$_2$ sphere and a 15\,nm VO$_2$ thin film on a TiO$_2$ (rutile) substrate across the MIT using a home-built scanning thermal microscopy setup.

\begin{figure}
\includegraphics[width=1.0\columnwidth,angle=0]{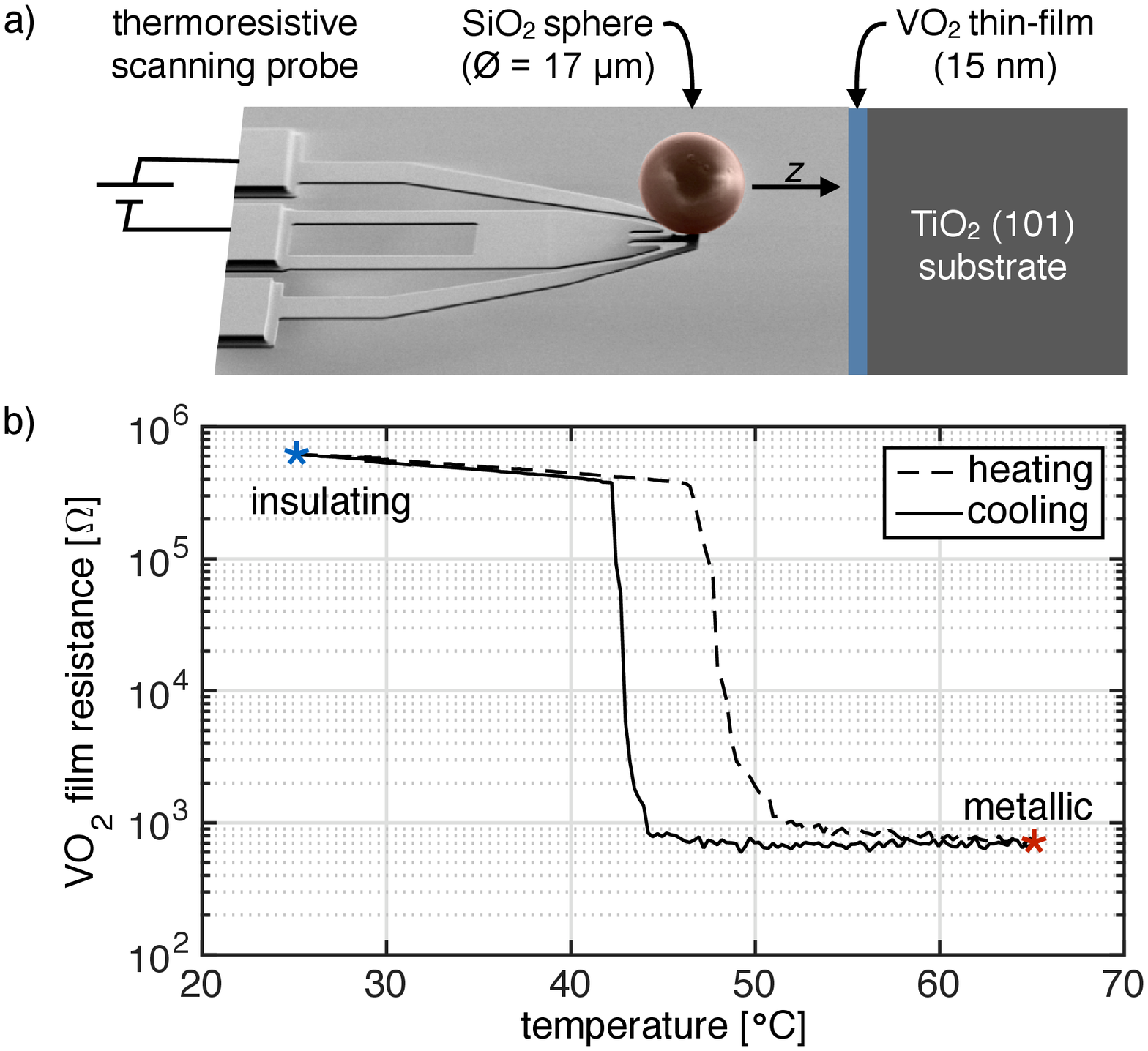}
\caption{a) Experimental arrangement including a scanning electron micrograph (artificially colored) of the cantilevered heater/sensor used in this study. \newline
b) Resistance versus temperature measurement of the VO$_2$ film used in this study. The temperatures set for the heat transport measurements are indicated with asterixes. \newline}
\label{FigScheme}
\end{figure}

A thin film of VO$_2$ was deposited on a single crystal TiO$_2$ (101) rutile substrate using pulsed laser deposition from a VO$_2$ target under a oxygen pressure of 10\,mTorr and a substrate temperature of 400\,$^\circ$C. Details of the growth of the VO$_2$ by pulsed laser deposition are described by Jeong et al. \cite{Jeong2015}. The resulting film is c-axis oriented with a thickness of 15$\pm$1\,nm as determined using ellipsometry.

Fig.\,\ref{FigScheme}b) shows the electrical resistance of the VO$_2$ sample as a function of temperature measured with two probes contacting the film at 450\,$\mu$m distance. The MIT occurs relatively sharply at around 46\,$^\circ$C with a hysteresis of about 6\,$^\circ$C. For the near field measurements, temperatures of 25\,$^\circ$C and 65\,$^\circ$C were chosen for the insulating and metallic state, respectively.

For the thermal transport measurements a vacuum-based ($10^{-7}$\,mbar) scanning thermal microscope \cite{Menges2013} was used with microfabricated cantilevered silicon thermal probes \cite{Drechsler2003}. Borosilicate glass microspheres (by Duke Standards) were attached to the sensors and served as one of the two surfaces for the near-field transport measurements. The data set shown below used a sphere with a nominal and measured diameter (by scanning electron microscopy) of 18$\pm$1 and 17\,$\mu$m, respectively.
The surface roughness of the sphere is small enough to regard it as smooth in the analysis. After attaching the microsphere to the cantilevers using micro-manipulators, it was annealed at $\approx$300\,$^\circ$C in-situ to remove moisture and possible contaminants.

The VO$_2$ sample was mounted on a Peltier stage to vary the sample temperature. The setup is situated in IBM's Noisefree Labs \cite{Loertscher2013}. As shown in Fig.\,\ref{FigScheme}a), the thermal cantilever, comprising a resistive sensor/heater element, is  approximately right angled with respect to the VO$_2$ surface to combine the optical deflection sensing of the cantilever with a relatively large mechanical stiffness in direction of the VO$_2$ surface. In contrast to previous experiments, we only use the optical deflection signal to determine the mechanical contact point between the glass sphere and the VO$_2$ film surface (with an accuracy of about 5\,nm) while the thermal measurements are performed using the integrated resistive sensor. Care was taken to ensure that possible electrostatic interactions between the VO$_2$ sample and the integrated silicon resistive sensor not cause artifacts on the thermal sensor signal. Details on calibration of the thermal probe sensor and the extraction of thermal conductance values from variations of the measured electrical resistance can be found elsewhere \cite{Menges2012}.

Experiments with six sets of temperatures were chosen, 25\,$^\circ$C and 65\,$^\circ$C for the VO$_2$ sample and 135\,$^\circ$C, 156\,$^\circ$C and 184\,$^\circ$C for the glass sphere. After allowing tip and VO$_2$ film to equilibrate to their respective temperatures, data was taken as a function of piezo displacement with a sampling time of 20\,ms per 0.5\,nm displacement such that a thermal steady state could be assumed at any time. The temperature-dependent electrical resistance of the silicon heater/sensor was used to determine its temperature and dissipated power. Together with the thermal resistance of the cantilevered structure, a net heat flux was extracted (see Menges et al. \cite{Menges2012} for details on the procedure). With the sphere attached, we reached a 10\,nW-sensitivity in a the measurement bandwidth used in this experiment (50\,Hz acquisition rate and averaging of 100 data sets) at an accuracy of 20\% \cite{Menges2013}. For increasingly large sphere-surface separations, the heat flux asymptotically reaches a constant value, $\dot{Q}_{offset}$. This value was determined for each data set and subtracted from the data to obtain the near-field contribution only. For analysis, 100 data traces of heat flux versus distance were averaged for each set of temperatures.

\begin{figure*}
\includegraphics[width=1\columnwidth,angle=0]{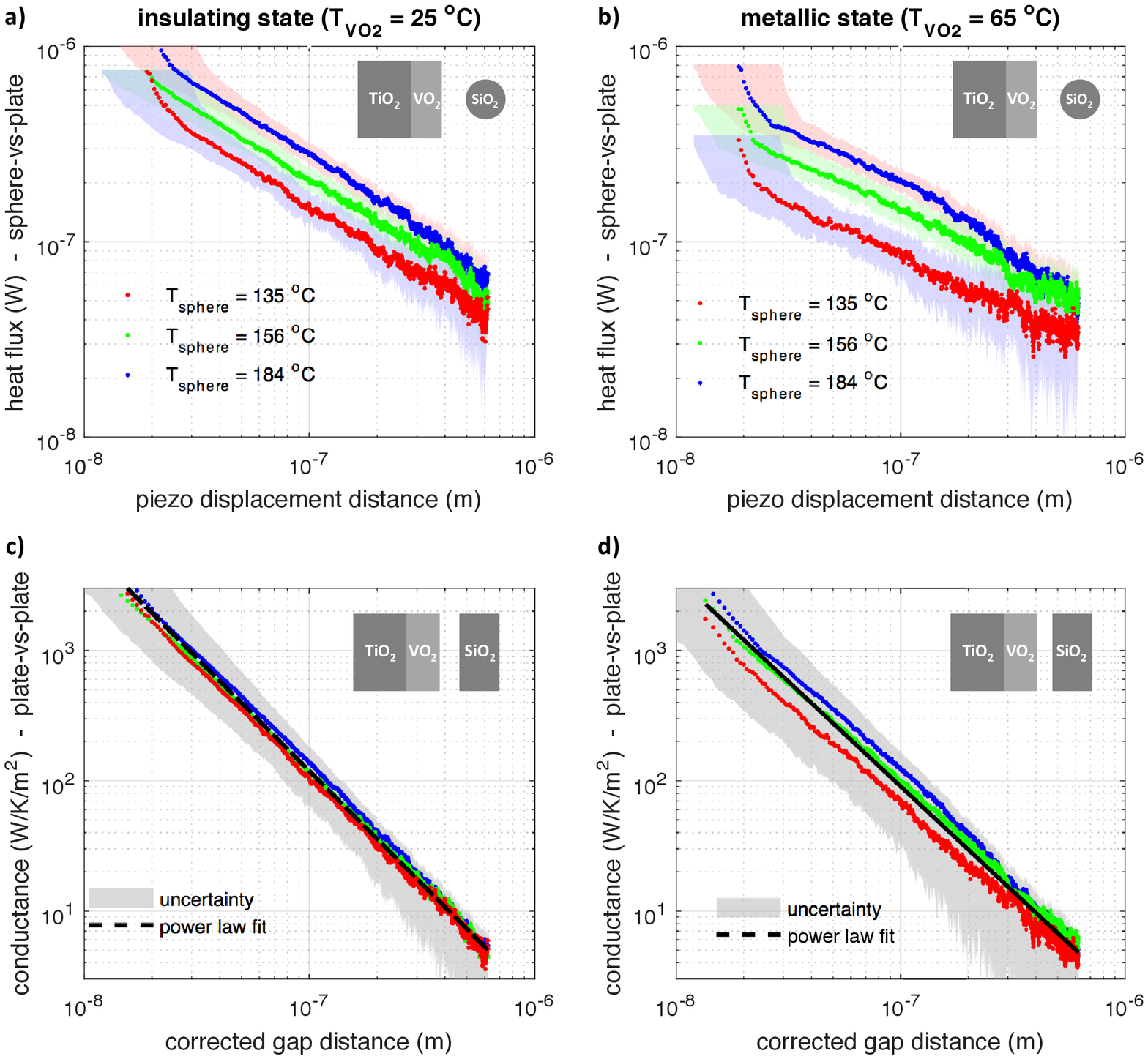}
\caption{(a,b) Experimentally measured heat flux between the sphere and the sample with VO$_2$ in the insulating state (a) and (b) metallic state as a function of piezo displacement. \newline Deduced thermal conductance vs. distance for the insulating (c) and (d) metallic state. 
The data in (c) and (d) were calculated from (a) and (b), respectively, by correcting for geometry, spring constant and temperature bias as described in the text. The data sets corresponding to the three sphere temperatures were fitted with a single power law function (dashed black line).}
\label{FigResults}
\end{figure*}

Fig.\,\ref{FigResults}a) and b) show the experimental heat flux data as a function of distance for the insulating and metallic film state, respectively, at different sample temperatures. The two main uncertainties are the constant $\dot{Q}_{offset}$ and the absolute distance. The respective uncertainties are indicated with color shading. They result in a change from the slope in the double logarithmic plot at the lowest and largest distances respectively.

To analyze the data further, we first calculate the conductance values from the heat flux. Near field thermal radiation between two bodies at respective temperatures $T_1$ and $T_2$ generally scale in proportion to $T_1^m - T_2^m$ with m$\le$2. Therefore, the temperature difference in our experiments may be large to apply linear response approximations. Let's assume a $T^2$ dependence to be verified by intercomparing the 3 data sets (see below). Then, by dividing the heat flux by $(T_1 - T_2) + (T_1-T_2)^2/(2T_2)$ instead of $(T_1 - T_2)$ we can recover the expression for conductance of the linear response regime under the assumption of the $T^2$ scaling. Together with the rescaling to a flat-vs.-flat geometry (see below) this makes the data comparable to theoretical predictions.

The attractive forces between glass sphere and VO$_2$ surface lead to an attractive force bending the cantilever and a resulting reduced distance between sphere and surface, as recognized in the data for distances below 30\,nm. The effect, however, can be easily subtracted using standard expressions for the force and using the distance between snap-in and zero deflection of the cantilever as a measure of the attractive forces.

Predictions of near field thermal transport as a function of separation are typically well described using power laws with a single exponent (at sufficiently small distances) on the order of $n=2$, depending on temperature and material.
The thermal conductance per unit area $g$ as a function of distance $z$ is then $g(z) = b\,z^{-n}$ with suitable constants $b$ and $n$. The effect of the sphere-plate geometry leads to a conductance of $G = 2\pi R b z^{-n+1}/(n-1)$ with $R$ being the sphere's radius. This approximation is valid for distances $z \ll R$ as in our case \cite{Song2015, Narayanaswamy2008, Otey2011}. Note that the transformation from $G$ to $g$ increases the exponent by one and thereby the dynamic range of the data.

Combining the three steps (rescaling the curves to the VO$_2$ temperatures, translating to a plane-plane geometry, and correcting for cantilever bending) allows us to replot the data in a form better suitable for comparison with theory. Fig.\,\ref{FigResults}c) and d) show the resulting conductance per unit area for the two VO$_2$ temperatures. The data for the two different sphere temperatures coincide well within the given uncertainties and demonstrate the validity of the $T^2$ temperature dependence. Power laws with a single exponent can be well fitted to the distance dependent data with exponents of $n=1.75\pm 0.25$ and $n=1.6\pm 0.25$ for the insulating and metallic state, respectively. As mentioned above, the uncertainties are predominantly due to errors in the offsets.

\begin{figure}
\includegraphics[width=1\columnwidth,angle=0]{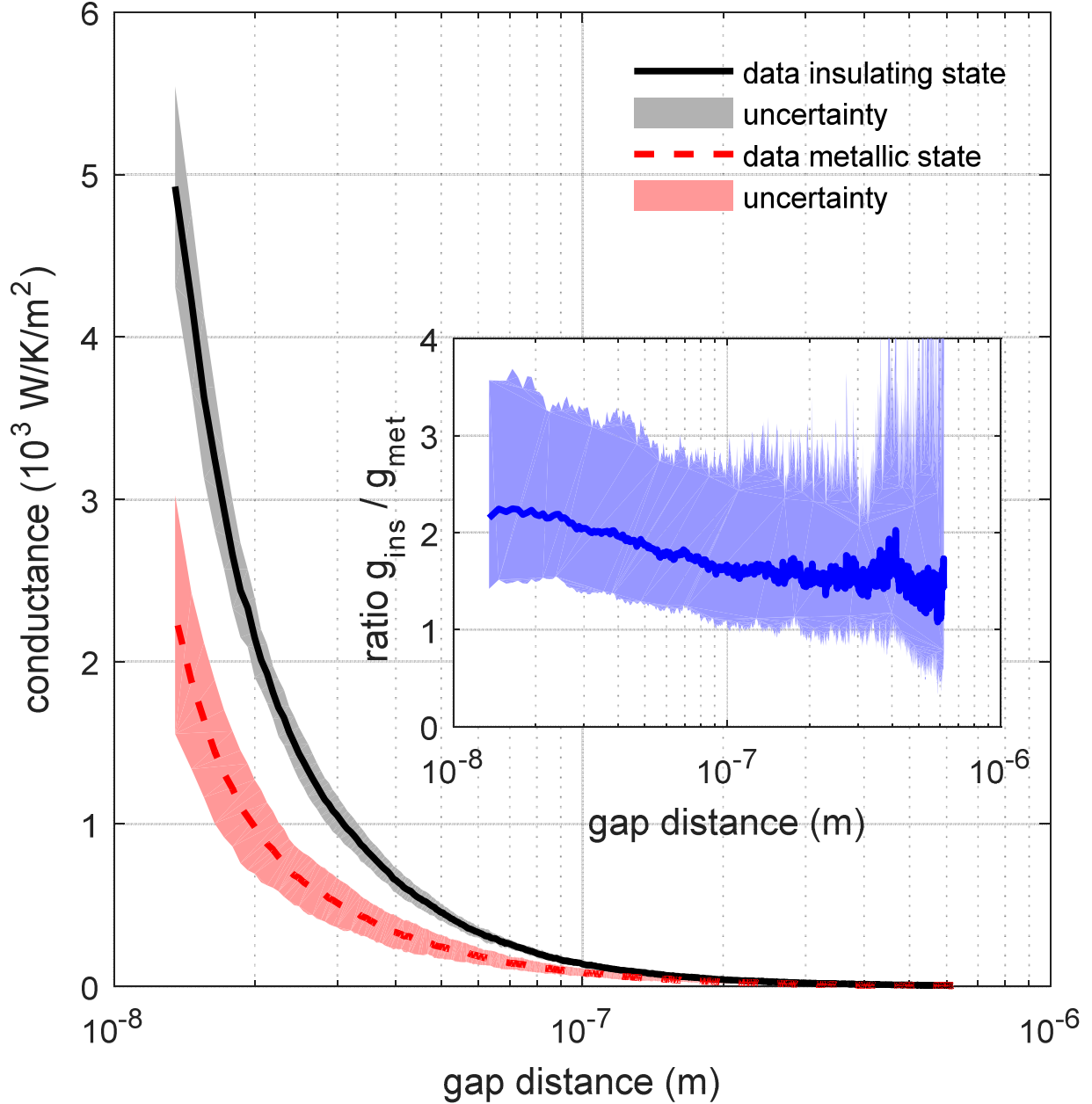}
\caption{Normalized thermal conductance vs. displacement and ratio $g_{ins}$/$g_{met}$ (inset) between the insulating and metallic state. The curves were taken from averaging the three data sets in Fig.\,\ref{FigResults}c) and d), respectively, after rescaling to a common temperature of 46\,$^\circ$C.}
\label{FigRatio}
\end{figure}

In view of possible applications as a thermal switch, the signal ratio between the two states was calculated. To allow a clear distinction between the two states of VO$_2$ temperatures well above and below the MIT were chosen. For a meaningful determination of the conductance contrast achievable, we rescaled the data to the same temperature in the hysteresis region of the MIT at 46\,$^\circ$C using the aforementioned $T^2$ dependence. The resulting conductance is plotted in Fig.\,\ref{FigRatio}. From the conductance in the metallic state $g_{met}$ and in the insulating state $g_{ins}$, we calculate a  ratio of $g_{ins}/g_{met}$, as shown in the inset of Fig.\,\ref{FigRatio}.

To discuss, we turn to existing predictions of the thermal near field transport and its contrast for VO$_2$ systems. We note, that the VO$_2$ film thickness of 15\,nm is too thin to treat it as a half space. There are no predictions available for such thin films on a rutile substrate. However, calculations of VO$_2$ thin films on a sapphire substrate \cite{Zwol2012} and on a SiO$_2$ substrate \cite{Yang2015} show a strongly reduced contrast between metallic and insulating state caused by the contributions from the substrates. Like sapphire and SiO$_2$, our TiO$_2$ (101 oriented, rutile) substrate also supports SPPs in the relevant frequency range and therefore we expect similar behavior.

As for the exponent $n$ in the insulating state, we note that the thin film geometry of VO$_2$ on sapphire is expected to have an exponent of $n=2$, similar to the case of two VO$_2$ half spaces ($n \approx 2$ in Yang et al.\cite{Yang2015}, $n \approx 1.7$ in van Zwol et al. \cite{Zwol2011}) and facing half spaces of VO$_2$ and SiO$_2$ ($n \approx 1.5$ in van Zwol et al. \cite{Zwol2011}). This is in good agreement to our exponent of $n=1.75\pm0.25$. For the conducting state the comparison is more difficult, because the contribution of interaction of the rutile substrate to the SiO$_2$ half space probably masks the metallic state of VO$_2$.

The measured magnitude of the heat transfer in the insulating state is smaller than predicted\cite{Zwol2011,Yang2015} for a pure VO$_2$-VO$_2$ system by a factor of 2 to 3 and about as large as predicted \cite{Zwol2011} for facing half spaces of VO$_2$ and SiO$_2$. Furthermore, the measured conductance in the insulating state is comparable to the 50\,nm-film calculation on sapphire \cite{Zwol2012}. All these observations are plausible considering a significant contribution from the rutile substrate in our experiments and the low thickness of VO$_2$ in our case.

Our experiments show a contrast between the two phase states reaching a value of more than 2 even for a film of only 15\,nm thickness. The predictions mentioned above \cite{Zwol2011, Zwol2012, Yang2015} show a reduction from a factor of $\approx 100$ for bulk half spaces to values compatible with our experiments after reduction of the VO$_2$ thickness to a few tens of nanometers on different substrates. This indicates that thicker VO$_2$ films or the avoidance of substrates will increase the observable contrasts.

In summary, we have shown near field thermal transport data determined with a scanning thermal microscopy setup using resistive probes. Distance- and temperature-dependent measurements show power law behavior compatible with existing predictions, $z^n$ with $n=1.5$ to 2, and a temperature dependence compatible with $T^m$ with $m \approx 2$. A thermal conductivity contrast of up to 2.5 between the two states of VO$_2$ was observed.\\

Funding of the Swiss Federal Office of Energy under grant agreement SI/501093-01 is gratefully acknowledged.
Furthermore, we thank Marilyne Sousa for ellipsometry measurements, Martin Spieser for simulation of the sensor temperature, Ute Drechsler for fabrication of thermoresistive scanning probes, Meinrad Tschudy, Siegfried Karg, the BRNC team and Walter Riess for support. We thank P.-O. Chapuis and Hidekazu Tanaka for discussions.


\end{document}